\documentstyle[12pt,a4]{article}
\newcommand{\pain}{\indent}

\newcommand{\vareps}{\varepsilon}
\newcommand{\eqref}[1]{$(\ref{#1})$}
\begin{document}
% ***** TITLE ****************************************************
\begin{titlepage}
\hfill solv-int/9511001\\
\vspace{40pt}

\begin{center}
  \begin{Large}
    \noindent{\bf Discrete soliton equations and}\\
    {\bf convergence acceleration algorithms}\\
  \end{Large}

\vspace{25pt}

\noindent
\begin{large}
A. {\sc Nagai}$^{\ddag}$ and J. {\sc Satsuma}
\end{large}

\vspace{18pt}
\begin{small}
\noindent
{\it Department of Mathematical Sciences, University of Tokyo,\\
Komaba 3-8-1, Meguro-ku, Tokyo 153, Japan}
\end{small}
\end{center}
% ***** ABSTRACT *************************************************
\vfill\vfill
\hrule
\bigskip
\begin{abstract}
Some of the well-known convergence acceleration algorithms, when
viewed as two-variable difference equations, are equivalent to
discrete soliton equations. It is shown that the $\eta-$algorithm is
nothing but the discrete KdV equation. In addition, one
generalized version of the $\rho-$algorithm is considered to be
integrable discretization of the cylindrical KdV equation.
\end{abstract}
\bigskip
% ***** FOOTNOTE *************************************************
\hrule
\medskip\noindent
$^{\ddag}$ E-mail: nagai@sat.t.u-tokyo.ac.jp

\medskip
to appear in Phys. Lett. A
% ================================================================
\end{titlepage}
\section{Introduction}

\pain Recently, it has been claimed that good algorithms in the field of
numerical analysis play important roles in the theory of nonlinear
integrable systems. In 1982, Symes~\cite{Symes} pointed out that one
step in the QR algorithm, which is the most popular method to solve
matrix eigenvalue problems, is equivalent to time evolution of the finite
nonperiodic Toda equation. In 1992, Hirota, Tsujimoto, and
Imai~\cite{Hirotarims} showed that the LR algorithm, which is another
successful tool to find eigenvalues of a given matrix, is no other
than the time-discrete Toda equation. In 1993, Papageorgiou, Grammaticos,
and Ramani~\cite{Papa} showed that a well-known convergence
acceleration scheme, the $\vareps-$algorithm, is nothing but the
discrete potential KdV equation.

Our main interest in this paper is on the convergence
acceleration algorithms. Let $\{S_m\}$ be a sequence of numbers which
converges to $S_\infty$. In order to find $S_\infty$ by direct
calculation, we often need a large amount of data. In such cases we
transform the original sequence $\{S_m\}$ into another sequence
$\{T_m\}$ instead of calculating directly. If $\{T_m\}$
converges to $S_\infty$ faster than $\{S_m\}$, that is
\begin{equation}
  \lim_{m \rightarrow \infty} \frac{T_m - S_\infty}{S_m - S_\infty}=0,
\end{equation}
we say that the transformation $T : \{S_m\} \rightarrow \{T_m\}$
{\it accelerates the convergence} of the sequence $\{S_m\}$.
We now have many convergence acceleration algorithms. Among them, we
here focus our attention on the $\eta-$,
$\vareps-$, and $\rho-$algorithms [4-6]. We clarify that
there is a strong tie between these algorithms and discrete soliton
equations.

In \S 2, we show that Bauer's $\eta-$algorithm is considered
to be the discrete KdV equation in ref.~\cite{Hirota1}.
We also look over the result by Papageorgiou et al., the equivalence
between Wynn's $\vareps-$algorithm and the discrete potential KdV
equation. In \S 3, we introduce a different type of algorithm, Wynn's
$\rho-$algorithm. In spite of its similarity with the
$\vareps-$algorithm, it possesses noticeably different characteristics
not only as a convergence accelerator but also as a discrete soliton
equation. We show that the $\rho-$algorithm relates with the
cylindrical KdV equation~\cite{Maxon},
\begin{equation}
u_t + 6 u u_x + u_{xxx} + \frac{1}{2t} u = 0.
  \label{cylind}
\end{equation}
Concluding remarks are given in \S 4.

\section{The $\eta-$algorithm and the $\vareps-$algorithm}

\pain In this section we show that Bauer's $\eta-$algorithm~\cite{Bauer1},
which is one of the famous convergence acceleration algorithms, is
equivalent to the discrete KdV equation. Let initial values
$\eta_0^{(m)}$ and $\eta_1^{(m)}$ be
\begin{equation}
\eta_0^{(m)} = \infty, \
\eta_1^{(m)} = c_m \equiv \Delta S_{m-1}, \ (m=0,1,2,\ldots), \
S_{-1} \equiv 0,
  \label{eta1}
\end{equation}
where $\Delta$ is the forward difference operator given by $\Delta a_k =
a_{k+1} - a_k$.
Then all the other elements are calculated from the following recurrence
relations called the $\eta-$algorithm;
\begin{equation}
\left\{
\begin{array}{rcl}
  \eta_{2n+1}^{(m)} + \eta_{2n}^{(m)} &=&
  \eta_{2n}^{(m+1)} + \eta_{2n-1}^{(m+1)} \\
  \displaystyle{\frac{1}{\eta_{2n+2}^{(m)}} +
  \frac{1}{\eta_{2n+1}^{(m)}}} &=&
  \displaystyle{\frac{1}{\eta_{2n+1}^{(m+1)}} +
\frac{1}{\eta_{2n}^{(m+1)}}}
\end{array}
\right. \ ({\rm rhombus \ rules}).
  \label{eta}
\end{equation}
This defines a transformation of a given series $c_m =
\eta_1^{(m)}, m = 0,1,2,\ldots$ to a new series
$c_n' = \eta_n^{(0)}, n = 1,2,\ldots$ such that $\sum_{n=1}^\infty
c_n'$ converges more rapidly to the same limit $S_\infty$. The
quantities $\eta_n^{(m)}$ are given by the following ratios of Hankel
determinants;
\begin{eqnarray}
  \eta_{2n+1}^{(m)} &=&
\frac{\left|
\begin{array}{ccc}
c_{m} & \cdots & c_{m+n} \\
\vdots &  & \vdots \\
c_{m+n} & \cdots & c_{m+2n}
\end{array} \right| \cdot
\left|
\begin{array}{ccc}
c_{m+1} & \cdots & c_{m+n} \\
\vdots &  & \vdots \\
c_{m+n} & \cdots & c_{m+2n-1}
\end{array} \right|}
{\left|
\begin{array}{ccc}
\Delta c_{m} & \cdots & \Delta c_{m+n-1} \\
\vdots &  & \vdots \\
\Delta c_{m+n-1} & \cdots & \Delta c_{m+2n-2}
\end{array} \right| \cdot
\left|
\begin{array}{ccc}
\Delta c_{m+1} & \cdots & \Delta c_{m+n} \\
\vdots &  & \vdots \\
\Delta c_{m+n} & \cdots & \Delta c_{m+2n-1}
\end{array} \right|}, \label{etao}\\
\nonumber \\
  \eta_{2n+2}^{(m)} &=&
\frac{\left|
\begin{array}{ccc}
c_{m} & \cdots & c_{m+n} \\
\vdots &  & \vdots \\
c_{m+n} & \cdots & c_{m+2n}
\end{array} \right| \cdot
\left|
\begin{array}{ccc}
c_{m+1} & \cdots & c_{m+n+1} \\
\vdots &  & \vdots \\
c_{m+n+1} & \cdots & c_{m+2n+1}
\end{array} \right|}
{\left|
\begin{array}{ccc}
\Delta c_{m} & \cdots & \Delta c_{m+n} \\
\vdots &  & \vdots \\
\Delta c_{m+n} & \cdots & \Delta c_{m+2n}
\end{array} \right| \cdot
\left|
\begin{array}{ccc}
\Delta c_{m+1} & \cdots & \Delta c_{m+n} \\
\vdots &  & \vdots \\
\Delta c_{m+n} & \cdots & \Delta c_{m+2n-1}
\end{array} \right|} \label{etae}.
\end{eqnarray}

If we introduce dependent variable transformations,
\begin{equation}
X_{2n}^{(m)} = \frac{1}{\eta_{2n}^{(m)}}, \
X_{2n-1}^{(m)} = \eta_{2n-1}^{(m)},
\end{equation}
the $\eta-$algorithm~\eqref{eta} is rewritten as
\begin{equation}
X_{n+1}^{(m)} - X_{n-1}^{(m+1)} =
\frac{1}{X_{n}^{(m+1)}} - \frac{1}{X_{n}^{(m)}},
  \label{dKdV}
\end{equation}
{
which is the discrete KdV equation.
Let us replace variables $n$ and $m$ by
\begin{equation}
n = \frac{t}{\epsilon^3},\ m-\frac{1}{2} = \frac{x}{\epsilon} -
\frac{ct}{\epsilon^3},
\end{equation}
respectively and rewrite $X_n^{(m)}$ as $p + \epsilon^2
u(x-\epsilon/2,t)$, where $\epsilon$ is a small parameter and $c$, $p$
are finite constants related by
\begin{equation}
  1 - 2c = \frac{1}{p^2}.
\end{equation}
Then eq.~\eqref{dKdV} becomes
\begin{equation}
\epsilon^2 u(x-\frac{\epsilon}{2}+c\epsilon,t+\epsilon^3)
- \epsilon^2 u(x+\frac{\epsilon}{2}-c\epsilon,t-\epsilon^3)
= \frac{1}{p + \epsilon^2 u(x+\frac{\epsilon}{2},t)} -
\frac{1}{p + \epsilon^2 u(x-\frac{\epsilon}{2},t)}.
  \label{dKdV2}
\end{equation}
If we take the small limit of $\epsilon$, eq.~\eqref{dKdV2}
yields the KdV equation~\cite{Hirota1},
\begin{equation}
u_t - \frac{1}{p^3} u u_x
+ \frac{1}{48p^2}(1-\frac{1}{p^4}) u_{xxx} = 0
  \label{KdV}
\end{equation}
at the order of $\vareps^5$.
}

The discrete KdV eq.~\eqref{dKdV} is rewritten as
\begin{eqnarray}
\{ \tau(n+2,m-1)\tau(n-1,m) + \tau(n+1,m)\tau(n,m-1) \} \tau(n-1,m+1)
\nonumber \\
=  \{ \tau(n-2,m+1)\tau(n+1,m) + \tau(n-1,m)\tau(n,m+1) \} \tau(n+1,m-1),
  \label{tridKdV}
\end{eqnarray}
through a dependent variable transformation,
\begin{equation}
  X_{n}^{(m)} = \frac{\tau(n+1,m-1)\tau(n-1,m)}{\tau(n,m-1)\tau(n,m)}.
\label{dvt-dKdV}
\end{equation}
Subtracting $\tau(n-1,m+1)\tau(n+1,m-1)\tau(n,m)$ from both sides of
eq.~\eqref{tridKdV}, we obtain the bilinear form of the discrete KdV
equation,
\begin{equation}
\tau(n+2,m-1)\tau(n-1,m) + \tau(n+1,m)\tau(n,m-1)
- \tau(n+1,m-1)\tau(n,m) = 0.
  \label{bildKdV}
\end{equation}
It is noted that the solutions~\eqref{etao} and \eqref{etae} are
recovered by putting
\begin{eqnarray}
  \tau(2n,m) &\equiv& \left|
\begin{array}{ccc}
c_{m+1} & \cdots & c_{m+n} \\
\vdots &  & \vdots \\
c_{m+n} & \cdots & c_{m+2n-1}
\end{array} \right|,\label{tau-even}\\
\nonumber \\
  \tau(2n+1,m) &\equiv& \left|
\begin{array}{ccc}
\Delta c_{m+1} & \cdots & \Delta c_{m+n} \\
\vdots &  & \vdots \\
\Delta c_{m+n} & \cdots & \Delta c_{m+2n-1}
\end{array} \right| \label{tau-odd}
\end{eqnarray}
in eq.~\eqref{dvt-dKdV}.

    Next, following the result by Papageorgiou et al., we briefly
review the equivalence between the $\vareps-$algorithm and the
discrete potential KdV equation. The $\vareps-$algorithm originates
with Shanks~\cite{Shanks} and Wynn~\cite{Wynn1}. Define
$\vareps_0^{(m)}$ and $\vareps_1^{(m)}$ by
\begin{equation}
\vareps_0^{(m)} = 0, \ \vareps_1^{(m)} = S_m \ (m=0,1,2,\ldots).
\end{equation}
Then all the other quantities obey the following rhombus
rule called the $\vareps-$algorithm;
\begin{equation}
    (\vareps_{n+1}^{(m)} - \vareps_{n-1}^{(m+1)})
  (\vareps_{n}^{(m+1)} - \vareps_{n}^{(m)}) = 1.
\label{eps}
\end{equation}
According as $n$ becomes large, $\vareps_{2n+1}^{(m)}$ converges more
rapidly to $S_\infty$ as $m \rightarrow \infty$. On the other hand,
$\vareps_{2n}^{(m)}$ diverges as $m \rightarrow \infty$. This fact
reminds us of the idea of the singularity confinement~\cite{Gram},
since a singularity at $(2n,m)$ is confined and $\vareps_{2n+1}^{(m)}$
converges to the same limit as the original sequence
$\vareps_1^{(m)}$.

It has been shown that the $\vareps-$algorithm~\eqref{eps} is regarded
as the discrete potential KdV equation. The quantities
$\vareps_n^{(m)}$ are also given by the following ratios of Hankel
determinants; %\clearpage
\begin{eqnarray}
  \vareps_{2n+1}^{(m)}
&=& \frac{\left|
\begin{array}{cccc}
S_m & S_{m+1} & \cdots & S_{m+n} \\
S_{m+1} & S_{m+2} & \cdots & S_{m+n+1} \\
\vdots & \vdots & & \vdots \\
S_{m+n} & S_{m+n+1} & \cdots & S_{m+2n}
\end{array} \right|}
{\left|
\begin{array}{cccc}
\Delta^2 S_m & \Delta^2 S_{m+1} & \cdots & \Delta^2 S_{m+n-1} \\
\Delta^2 S_{m+1} & \Delta^2 S_{m+2} & \cdots & \Delta^2 S_{m+n} \\
\vdots & \vdots & & \vdots \\
\Delta^2 S_{m+n-1} & \Delta^2 S_{m+n} & \cdots & \Delta^2 S_{m+2n-2}
\end{array} \right|},
\label{Shanks}\\
\nonumber \\
  \vareps_{2n+2}^{(m)}
&=& \frac{\left|
\begin{array}{cccc}
\Delta^3 S_m & \Delta^3 S_{m+1} & \cdots & \Delta^3 S_{m+n-1} \\
\Delta^3 S_{m+1} & \Delta^3 S_{m+2} & \cdots & \Delta^3 S_{m+n} \\
\vdots & \vdots & & \vdots \\
\Delta^3 S_{m+n-1} & \Delta^3 S_{m+n} & \cdots & \Delta^3 S_{m+2n-2}
\end{array} \right|}
{\left|
\begin{array}{cccc}
\Delta S_m & \Delta S_{m+1} & \cdots & \Delta S_{m+n} \\
\Delta S_{m+1} & \Delta S_{m+2} & \cdots & \Delta S_{m+n+1} \\
\vdots & \vdots & & \vdots \\
\Delta S_{m+n} & \Delta S_{m+n+1} & \cdots & \Delta S_{m+2n}
\end{array} \right|} .
\end{eqnarray}
Equation~\eqref{Shanks} is called the Shanks transformation~\cite{Shanks}.
Substitution of $n=1$ in eq.~\eqref{Shanks} gives the well-known
Aitken acceleration algorithm.

   We have so far seen that the $\eta-$ and the $\vareps-$algorithms
are interpreted as the discrete KdV and the discrete potential KdV
equations, respectively. Therefore, these two algorithms are the same
in their performance as convergence acceleration algorithms. This
equivalence can also be understood from the fact~\cite{Bauer1} that
the quantities $\eta_n^{(m)}$ and $\vareps_n^{(m)}$ are related by
\begin{equation}
\eta_{2n}^{(m)} = \vareps_{2n+1}^{(m-1)} - \vareps_{2n-1}^{(m)}, \
 \eta_{2n+1}^{(m)} = \vareps_{2n+1}^{(m)} - \vareps_{2n+1}^{(m-1)}.
  \label{eta-eps}
\end{equation}

\section{The $\rho-$algorithm}

\pain The $\rho-$algorithm is traced back to Thiele's rational
interpolation~\cite{Thiele}. It was first used as a convergence
accelerator by Wynn~\cite{Wynn2}. The initial values of the algorithm
are given by
\begin{equation}
  \rho_0^{(m)} = 0, \ \rho_1^{(m)} = S_m \ (m=0,1,2,\ldots),
\end{equation}
and all the other elements fulfill the following rhombus rule;
\begin{equation}
(\rho_{n+1}^{(m)}-\rho_{n-1}^{(m+1)})(\rho_n^{(m+1)}-\rho_n^{(m)}) = n.
  \label{rho}
\end{equation}
The $\rho-$algorithm is almost the same as the $\vareps-$algorithm
except that ``1'' in the right hand side of eq.~\eqref{eps} is
replaced by ``$n$'' in eq.~\eqref{rho}. This slight change, however,
yields considerable differences in various aspects between these two
algorithms.

   The first difference is in their performance. As one can find in
ref.~\cite{Smith}, the $\vareps-$algorithm accelerates exponentially
or alternatively decaying sequences, while the $\rho-$algorithm
does rationally decaying sequences.

The second difference is in their determinant expressions. The quantities
$\vareps_n^{(m)}$ are given by ratios of Hankel determinants, while the
quantities $\rho_n^{(m)}$ are given by~\cite{Thiele}
\begin{equation}
\rho^{(m)}_n = (-1)^{\left[\frac{n-1}{2}\right]}
\frac{\tilde\tau_n^{(m)}}{\tau_n^{(m)}},
  \label{rho2}
\end{equation}
where $[x]$ stands for the greatest integer less than or equal to $x$.
Moreover, the functions $\tau_n^{(m)}$ and $\tilde\tau_n^{(m)}$ are
expressed as the following double Casorati determinants;
\begin{eqnarray}
\tau_n^{(m)} &=& \left\{
\begin{array}{ll}
  u^{(m)}(k;k) & n = 2k, \\
  u^{(m)}(k+1;k) & n = 2k+1,
\end{array} \right.
  \label{rhof} \\
\tilde\tau_n^{(m)} &=& \left\{
\begin{array}{ll}
  u^{(m)}(k+1;k-1) & n = 2k, \\
  u^{(m)}(k;k+1) & n = 2k+1,
\end{array} \right.
  \label{rhog}
\end{eqnarray}
%\clearpage
where
\begin{eqnarray}
&& u^{(m)}(p;q) =
\det\left[
\begin{array}{ccccl}
1 & m & \cdots & m^{p-1} & | \\
1 & m+1 & \cdots & (m+1)^{p-1} & | \\
\vdots & \vdots & & \vdots & | \\
1 & m+p+q-1 & \cdots & (m+p+q-1)^{p-1} & |
\end{array}
\right.
\nonumber \\
%\nonumber \\
&& \left. \begin{array}{rcccc}
| & S_m & m S_m & \cdots & m^{q-1} S_m \\
| & S_{m+1} & (m+1) S_{m+1} & \cdots & (m+1)^{q-1} S_{m+1} \\
| & \vdots & \vdots & & \vdots \\
| & S_{m+p+q-1} & (m+p+q-1) S_{m+p+q-1} &
\cdots & (m+p+q-1)^{q-1} S_{m+p+q-1}
\end{array} \right] \label{taurho}.
\end{eqnarray}

   The third difference is in the relation with discrete soliton
equations. We have seen in the previous section that the
$\vareps-$algorithm is regarded as the discrete potential KdV
equation. Before discussing the relation of the $\rho-$algorithm with
soliton equations, let us survey the result by Papageorgiou et al.
again. They considered the most general form of the algorithm given by
\begin{equation}
    (x_{n+1}^{(m)} - x_{n-1}^{(m+1)})
  (x_{n}^{(m+1)} - x_{n}^{(m)}) = z_n^{(m)},
\label{grho}
\end{equation}
and applied the singularity confinement test to eq.~\eqref{grho}. As a
result, when $z_n^{(m)}$ is of the form,
\begin{equation}
  z_n^{(m)} = f(n+m) + g(m),
\end{equation}
eq.~\eqref{grho} passes the test and is expected to be integrable. If
we put
\begin{equation}
  f(x) = x, \ g(x)=-x,
\end{equation}
eq.~\eqref{grho} gives the $\rho-$algorithm~\eqref{rho}, which
indicates that there is a chance for eq.~\eqref{rho} to be some
discrete analogue of soliton equations. Instead of the
$\rho-$algorithm~\eqref{rho} itself, we consider the algorithm of
the following form;
\begin{equation}
    (\rho_{n+1}^{(m)} - \rho_{n-1}^{(m+1)})
  (\rho_{n}^{(m+1)} - \rho_{n}^{(m)}) = an + b(m+1),
  \label{crho}
\end{equation}
where $a$ and $b$ are constant. Employing a dependent variable
transformation,
\begin{equation}
  Y_n^{(m)} = \rho_n^{(m)} - \rho_n^{(m-1)},
\label{Yrho}
\end{equation}
we obtain
\begin{equation}
Y_{n+1}^{(m)} - Y_{n-1}^{(m+1)} = \frac{an+bm+b}{Y_{n}^{(m+1)}} -
\frac{an+bm}{Y_{n}^{(m)}}
  \label{lrho}
\end{equation}
from eq.~\eqref{crho}. Equation~\eqref{lrho} possesses a form similar
to the discrete KdV eq.~\eqref{dKdV}. However, the nonautonomous
property of eq.~\eqref{lrho} yields an essential difference in its
continuous limit.
Let us introduce new variables $t, x$ defined by
\begin{equation}
  \frac{t}{\epsilon^3} = an + bm, \
\frac{x}{\epsilon} = cn + m-\frac{1}{2},
\end{equation}
and rewrite $Y_n^{(m)}$ as $\epsilon^{-3/2} \sqrt{t} \left\{ p +
\epsilon^2 u(x-\epsilon/2,t)\right\}$,
where $\epsilon$ is a small parameter and $p, c$ are finite constants
satisfying
\begin{equation}
  p^2 = \frac{b}{2a-b}, \ c = \frac{1}{2}-\frac{1}{2p^2}
= \frac{b-a}{b}.
\label{pbc}
\end{equation}
Then eq.~\eqref{lrho} becomes
\begin{eqnarray}
&& \epsilon^2 u(x-\frac{\epsilon}{2}+c\epsilon,t+a\epsilon^3)
- \epsilon^2 u(x+\frac{\epsilon}{2}-c\epsilon,t+(b-a)\epsilon^3)
\nonumber \\
&-& \frac{1}{p + \epsilon^2 u(x+\frac{\epsilon}{2},t+b\epsilon^3)}
+\frac{1}{p + \epsilon^2 u(x-\frac{\epsilon}{2},t)} \nonumber \\
&+& \frac{\epsilon^3}{2t}\left[
a\left\{p+\epsilon^2 u(x-\frac{\epsilon}{2}+c\epsilon,t+a\epsilon^3)\right\}
\right. \nonumber \\
&& \quad + \ (a-b)\left\{p+\epsilon^2 u(x+\frac{\epsilon}{2}-c\epsilon,t+(b-a)
\epsilon^3)\right\} \nonumber \\
&& \quad - \ \left.\frac{b}{p + \epsilon^2
u(x+\frac{\epsilon}{2},t+b\epsilon^3)}
\right] = 0.  \label{crho3}
\end{eqnarray}
Taking the small limit of $\epsilon$, we have
\begin{equation}
(2a-b)u_t - \frac{1}{p^3}u u_x
+ \frac{1}{48p^2}(1-\frac{1}{p^4})u_{xxx} +
\frac{(2a-b)}{2t}u = 0
  \label{ckdv}
\end{equation}
at the order of $\epsilon^5$ from eq.~\eqref{crho3}. Since the
coefficient of $u_t$ is always twice as large as that of $u/t$,
eq.~\eqref{crho} is considered as one integrable discretization of the
cylindrical KdV equation. It is interesting to note that the
$\rho-$algorithm~\eqref{rho} is not exactly discretization of the
cylindrical KdV equation. This is because we have $p=0$ in
eq.~\eqref{pbc} and coefficients of $uu_x$ and $u_{xxx}$ become
infinite in the case of the $\rho-$algorithm.

The third difference can be understood clearly from a viewpoint of the
Hirota's formalism. Employing the same dependent variable
transformation as eq.~\eqref{Yrho}, we obtain
\begin{equation}
Y_{n+1}^{(m)} - Y_{n-1}^{(m+1)} = \frac{n}{Y_{n}^{(m+1)}} -
\frac{n}{Y_{n}^{(m)}}
  \label{llrho}
\end{equation}
from eq.~\eqref{rho}. Moreover, through the same dependent variable
transformation as eq.~\eqref{dvt-dKdV},
\begin{equation}
  Y_n^{(m)} = \frac{F(n+1,m-1)F(n-1,m)}{F(n,m-1)F(n,m)},
\end{equation}
eq.~\eqref{llrho} is rewritten as the following trilinear form;
\begin{equation}
\left|
\begin{array}{ccc}
  -F(n+2,m-1) & F(n+1,m-1) & n F(n,m-1) \\
  F(n+1,m) & 0 & F(n-1,m) \\
  -n F(n,m+1) & F(n-1,m+1) & F(n-2,m+1)
\end{array}
\right| = 0.
  \label{lrhotri}
\end{equation}
The functions $F(n,m)$ and $\tau^{(m)}_n$ in eq.~\eqref{rhof} are
related by
\begin{equation}
  F(n,m) = (-1)^{a(n)} \tau^{(m)}_n,
\end{equation}
where $a(n)$ satisfies
\begin{equation}
  a(n) \equiv a(n-2) + \left[\frac{n-2}{2}\right] \ ({\rm mod} \ 2), \
  a(0) = a(1) = 0.
\end{equation}
Because of nonautonomous property of eq.~\eqref{lrho}, there is no way
to derive a bilinear form with a single variable $F(n,m)$ from the
trilinear eq.~\eqref{lrhotri}. This fact reminds us of the similarity
constraint of the discrete KdV equation~\cite{Satsuma}. It should be
noted, however, that a pair of functions $\tau_n^{(m)}$ and
$\tilde\tau_n^{(m)}$ given by eqs.~\eqref{rhof} and \eqref{rhog}
satisfy bilinear equations,
\begin{eqnarray}
\tau^{(m)}_{n+1} \tau^{(m+1)}_{n-1} -
\tau^{(m)}_n \tilde\tau^{(m+1)}_n + \tau^{(m+1)}_n \tilde\tau^{(m)}_n = 0,
  \label{bil1} \\
\tau^{(m+1)}_{n-1} \tilde\tau^{(m)}_{n+1} +
\tau^{(m)}_{n+1} \tilde\tau^{(m+1)}_{n-1} -
n \tau^{(m)}_n \tau^{(m+1)}_n = 0,
  \label{bil2}
\end{eqnarray}
which are considered to be the Jacobi and the Pl\"{u}cker identities
for determinants, respectively.

\section{Concluding Remarks}

\pain We have seen that the $\eta-$algorithm and the
$\vareps-$algorithm are equivalent to the discrete KdV and the
discrete potential KdV equations, respectively and that their
performance as convergence acceleration algorithms is completely the
same. We have also shown that the one generalization of the
$\rho-$algorithm is considered as integrable discretization of the
cylindrical KdV equation. The $\vareps-$ and the $\rho-$ algorithms,
despite their apparent similarity, possess different properties both
as convergence accelerators and as discrete soliton equations. The
difference in performance of these two algorithms depends on their
different determinant expressions.

When we apply $\vareps-$ and $\rho-$algorithms to a convergent
sequence, odd terms converge to the same limit as the original
sequence though even terms diverge. This fact agrees with the idea of
the singularity confinement. It is a future problem to clarify how two
different notions, acceleration and integrability, are associated with
each other. In other words, we should consider whether we can
construct new convergence acceleration algorithms from the other
discrete soliton equations\footnote{Actually, Papageorgiou et al. have
proposed a new algorithm based on the discrete modified KdV equation.}
and what kind of equations the other algorithms correspond to. The
solution of these problems will shed a new light on the study of
discrete integrable systems and numerical analysis.


\begin{thebibliography}{99}
%
\bibitem{Symes}W. W. Symes, Physica {\bf 4}D (1982) 275.
%
\bibitem{Hirotarims}R. Hirota, S. Tsujimoto, and T. Imai, RIMS Kokyuroku
{\bf 822} (1992) 144.
%
\bibitem{Papa}V. Papageorgiou, B. Grammaticos, and A. Ramani,
Phys. Lett. A {\bf 179} (1993) 111.
%
\bibitem{Bauer1}F. L. Bauer, in: On Numerical Approximation,
ed. R. E. Langer (University of Wisconsin Press, Madison, 1959) p.361.
%
\bibitem{Wynn1}P. Wynn, Math. Tables Aids Comput. {\bf 10} (1956) 91.
%
\bibitem{Wynn2}P. Wynn, Proc. Cambridge Philos. Soc. {\bf 52} (1956) 663.
%
\bibitem{Hirota1}R. Hirota, J. Phys. Soc. Jpn. {\bf 43} (1977) 1424,
{\bf 50} (1981) 3785.
%
\bibitem{Maxon}S. Maxon and J. Viecelli, Phys. Fluids {\bf 17} (1974) 1614.
%
\bibitem{Shanks}D. Shanks, J. Math. Phys. {\bf 34} (1955) 1.
%
\bibitem{Gram}B. Grammaticos, A. Ramani, and V. Papageorgiou,
Phys. Rev. Lett. {\bf 67} (1991) 1825.
%
\bibitem{Thiele}T. N. Thiele, Interpolationsrechnung
(Taubner, Leibzig, 1909).
%
\bibitem{Smith}D. A. Smith and W. F. Ford, SIAM J. Numer. Anal. {\bf 16}
(1979) 223.
%
\bibitem{Satsuma}J. Satsuma, A. Ramani, and B. Grammaticos,
Phys. Lett. A 174 (1993) 387.
%
\bibitem{Brezinsky}C. Brezinski, Lecture notes in mathematics, Vol. 584,
Acc\'{e}l\'{e}ration de la convergence and analyse num\'{e}rique
(Springer, Berlin, 1977).
%
\end{thebibliography}
\end{document}